\documentclass{aa}
\usepackage[dvips]{graphics}
\begin{document}

   \thesaurus{6     % A&A Section 6: Formation, structure and evolution of stars
              (08.03.2;  %  Stars: chemically peculiar(ought to be helium class)
	       08.15.1;  % Stars: oscillations (including pulsations)
	       08.09.2 (LSS~3184);  % Stars: individual (LSS 3184)
               08.22.3)} % Stars: variables: other
   \title{Physical properties of the pulsating hydrogen-deficient star
\object{LSS~3184} (\object{BX~Cir})
\thanks{Based on observations obtained with the NASA/ESA
{\it Hubble Space Telescope}, which is operated by STScI for the
Association of Universities for Research in Astronomy, Inc., under NASA
contract NAS 5-26555. Based on observations obtained at the
Anglo-Australian Telescope, Coonabarabran, NSW, Australia.}}

   \author{V. M. Woolf
          \and
          C. S. Jeffery
          }

   \offprints{V. M. Woolf}

   \institute{Armagh Observatory, College Hill, Armagh, BT61~9DG,
		Northern Ireland \\
              email: vmw@star.arm.ac.uk, csj@star.arm.ac.uk
             }

   \date{Received 16 February 2000; accepted 3 April 2000}

   \titlerunning{Physical properties of LSS~3184}
   \maketitle

   \begin{abstract}

We report new determinations of the radius and mass of the pulsating
helium-rich, hydrogen-deficient star \object{LSS~3184} (\object{BX Cir})
using measurements of
radial velocity and angular radius throughout its pulsation cycle. Measurements
of radial velocity, and thus changes in stellar radius ($\Delta R_\star$),
were made using Anglo-Australian
Telescope echelle spectra.  {\it Hubble Space Telescope} ultraviolet spectra
and ground-based {\it BV} photometry were used to
find temperatures and fluxes throughout
the pulsation cycle.  The temperatures and fluxes were used to find the angular
radius of the star ($\alpha$).  The $\alpha$, $\Delta \alpha$, and $\Delta R$
values thus found were used to calculate the mean stellar radius
$\langle R_\star \rangle = 2.31 \pm 0.10 R_\odot $.
If we use the previously determined $\log g = 3.35 \pm 0.1$ for
\object{LSS~3184} and our
radius estimate, we find its mass to be $M_\star = 0.42 \pm 0.12 M_\odot$.  

   \end{abstract}

%
%________________________________________________________________

\section{Introduction}

Pulsations in stars provide tools for researchers in several fields of
astronomy.  They provide standard candles for measuring Galactic and
extragalactic distances and they provide methods for measuring stellar
parameters, even below the directly observable photosphere.  Pulsations in
hydrogen-deficient, helium-rich stars such as extreme helium, R Coronae
Borealis, and hydrogen-deficient carbon stars
have not been studied in as much detail as those in
stars with more `normal' chemical abundances.  While this is partly
understandable since most pulsating stars have normal compositions, studying
stars with little or no hydrogen is important to allow tests of pulsation
theory.

Extreme helium stars, as their name would imply, have weak or
non-existent hydrogen absorption lines and very large helium abundances
($\ga 99$ per cent). Two extreme helium
stars, \object{V652~Her} and \object{LSS~3184}, are known to pulsate.
Saio (\cite{s93}) showed that in extreme helium stars with temperatures around
$2 \times 10^5$~K, like \object{V652~Her}, the $\kappa$-mechanism caused by
iron-group (Z-bump) opacity can excite the observed pulsations.
Saio (\cite{s94}, \cite{s95}) predicted that \object{LSS~3184}
should pulsate because of its location in the
Z-bump instability finger.  Kilkenny \&\ Koen (\cite{kk95}) discovered
that \object{LSS~3184} shows photometric variations with a period of about
0.107~d.  The similarity of \object{V652~Her} and 
\object{LSS~3184} in temperature, surface gravity, and pulsation period implies
that they are very similar in other physical parameters.

Kilkenny et~al. (\cite{kk99}) have recently reported an observational analysis
of \object{LSS~3184}, including a determination
of its photometric period (0.1066~d).  In
addition, they used medium resolution ($\lambda / \Delta \lambda \approx 4000$)
spectra to measure radial velocity variations and to show that the
photometric  variability is caused by pulsations.  Drilling,
Jeffery, \& Heber (\cite{d98}) reported an analysis of \object{LSS~3184}
in which they found
$T_{\rm eff} = 23\,300 \pm 700$~K, $\log g = 3.35 \pm 0.1$ and
$n_{\rm H}/n_{\rm He} \le 0.00015$. Kilkenny et~al. (\cite{kk99}) report
that using the radius they determined and $\log g$ from
Drilling et~al. (\cite{d98}) gives a mass of
$0.15 M_\odot$, which is much smaller than $0.7 M_\odot$, the mass accepted for
\object{V652~Her}
(Lynas-Gray et~al. \cite{l84}), and small enough to imply that some input
parameter or procedure is in error.

In this paper we report an analysis of \object{LSS~3184}
using high resolution optical
spectra for radial velocity measurement and ultraviolet {\it Hubble Space
Telescope} ({\it HST}) spectra and ground-based
{\it BV} photometry for temperature, luminosity, and angular radius
measurement.  The new data provide better temperature and angular radius
estimates and a much cleaner radial velocity curve for the star, allowing a
more reliable estimate of its radius and mass than was possible previously.

\section{Observations and data reduction}

\begin{figure*}
\resizebox{\hsize}{!}{\rotatebox{270}{\includegraphics{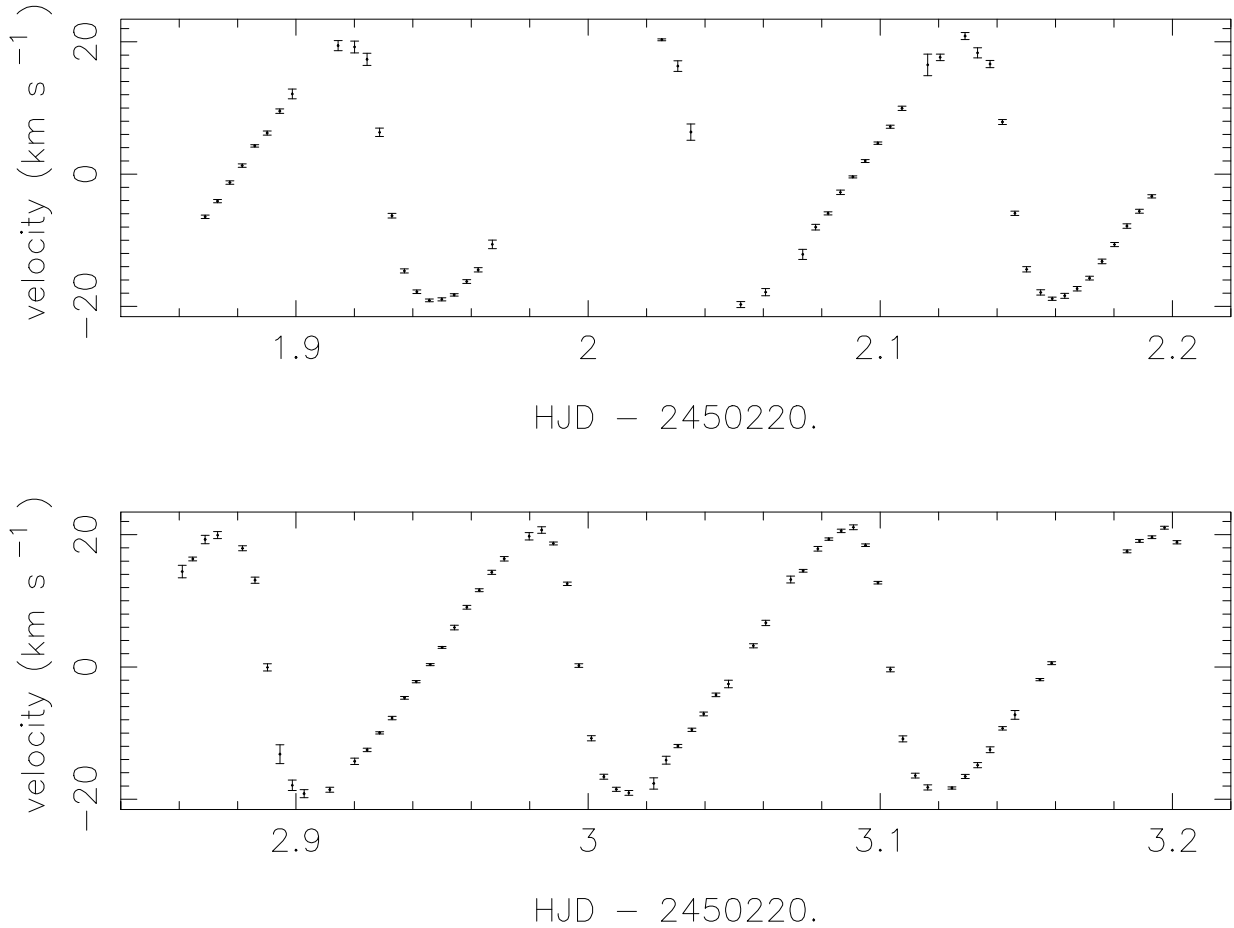}}}
\caption{Plot showing velocity and phase coverage for \object{LSS~3184}
spectroscopic observations, with error bars shown for velocity.
Note that the velocity here is not corrected for projection effects.}
\label{figa}
\end{figure*}

\subsection{AAT visible spectral observations}
Spectra of \object{LSS~3184} were obtained during
the nights of 1996 May 18 and 19 using the
University College London Echelle Spectrograph at the 3.9-m
Anglo-Australian Telescope.
Exposure times were between 4 and 5 minutes. Standard {\sc iraf} packages were
used for bias and flat field correction, reducing echelle orders to one
dimensional spectra, and applying the wavelength scale using thorium-argon
spectra.  As the wavelengths covered by adjacent orders overlapped, the spectra
covered the range 3850--5055~\AA\ completely.  The spectral resolution was
$\lambda / \Delta \lambda \approx 48\,000$.
Velocity corrections for Earth's motion were found for each exposure
using {\sc rvcorrect} and were applied using {\sc dopcor}.

\subsection{{\it Hubble Space Telescope} ultraviolet observations}
Ultraviolet spectra of \object{LSS~3184} were obtained using the Faint Object
Spectrograph of the {\it HST}.  Observations were made in RAPID mode using the
blue detector, the $0\farcs 86$ square aperture and grating G160L over three
orbits on 1997 February 7.  Data files reduced using the most recent
calibration files were downloaded from the {\it HST} archives on 1999 October
12. Individual exposures were made approximately 19.25 seconds apart.
The usable part of the spectra covered the range 1150--2500~\AA .  Based on the
ephemeris of Kilkenny et~al. (\cite{kk99}), observations during the three orbits
covered the phase ($\phi$) ranges from 0.3154--0.6099, 0.9327--0.2376, and
0.5602--0.8651, where maximum{\it V} magnitude is at $\phi = 0$ and again at
$\phi = 1$. Thus the star was observed by the {\it HST}
over 85.5 per cent of its pulsational cycle.

\section{Analysis}

\subsection{Radial velocity determinations}
The velocity shifts between spectra were measured using the cross correlation
package {\sc fxcor} in {\sc iraf}. For each measurement the velocities found
from 28 of the 35 orders of the echelle spectra
were used to find a weighted average velocity.
Weights for the averaging were the inverse of the velocity errors reported by
{\sc fxcor}.  The unused spectral orders either had no strong absorption lines
or had strong interstellar lines which did not allow a reliable stellar velocity
determination.

The procedure took two iterations.  In the
first iteration the best results were obtained by using the sum of all the
second night's \object{LSS~3184}
spectra as the cross correlation template for the first
night's spectra and vice versa.  The spectra from both nights were then shifted
by the velocities thus found and co-added to provide the template for the second
iteration.  The absorption lines in the second template were much sharper, as
the velocity smearing due to the stellar pulsations was effectively removed.
A third iteration was performed with the spectra shifted by the velocities found
in the second iteration before co-adding to make the template,
but the velocities
from the third iteration were effectively identical to those from the second.
The velocities from the second iteration, shifted so that the mean velocity is
zero, are shown in Fig.~\ref{figa}.  Error bars show the
weighted standard deviations of the velocity from the echelle orders.
Gaps are present in the data where clouds prevented observations.
From the the data in Fig.~\ref{figa}, we find that the peak-to-peak radial
%\begin{figure*}
%\resizebox{\hsize}{!}{\rotatebox{270}{\includegraphics{2034.f1}}}
%\caption{Plot showing velocity and phase coverage for \object{LSS~3184}
%spectroscopic observations, with error bars shown for velocity.
%Note that the velocity here is not corrected for projection effects.}
%\label{figa}
%\end{figure*}
velocity variation is $40.0\pm 0.3$~km~$\rm s^{-1}$, which is larger than
30~km~$\rm s^{-1}$, the value found by Kilkenny et~al. (\cite{kk99}).

\begin{figure}
\resizebox{\hsize}{!}{\rotatebox{270}{\includegraphics{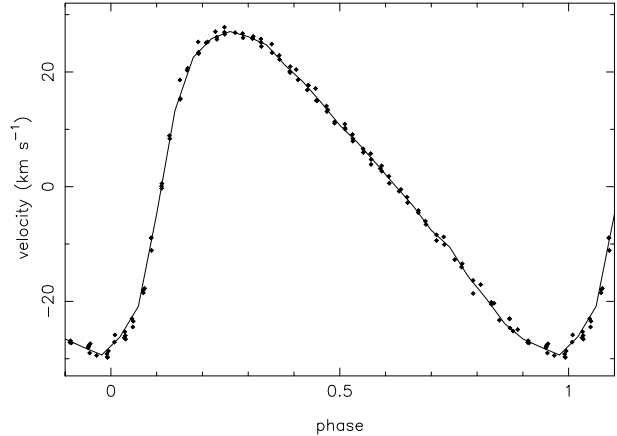}}}
\caption{Fit (solid line) to velocity data corrected for projection effects
(points). The data have been folded over by 0.2 cycles in Figs. \ref{figa2},
\ref{figc}, and \ref{figd}.}
\label{figa2}
\end{figure}

Velocity data from the two nights were phased to the pulsation cycle
using the ephemeris of Kilkenny et~al. (\cite{kk99}):
$T_0 = 2\,449\,477.4691 (\pm 0.0016)$, ${\rm Period} = 0.1065784
(\pm 0.0000005) {\rm d}$.  The measured radial velocities were multiplied by the
factor $-1.42$ to correct for projection effects and give the surface
velocity through the pulsation cycle in the stellar rest frame.
The projection
factor was chosen based on preliminary work by Monta\~n\'es Rodriguez
et~al. (\cite{m00}) (See also Albrow \&\ Cottrell \cite{a94}; Gautschy
\cite{g87}; and references therein).
We will discuss later the effects of choosing a different projection factor.
A smooth curve (high order polynomial)
was fit to the velocity data (Fig.~\ref{figa2}).  The velocity
values on this curve were used to calculate change in stellar radius
($\Delta R_\star$) and surface acceleration.
The phase bin centers for acceleration and $\Delta R_\star$ are
shifted by half a bin with respect to the velocity bins, i.e. the end of an
acceleration bin is the center of the next velocity bin and the end of a
velocity bin is the center of the next $\Delta R_\star$ bin.

\subsection{Temperature determination}
To find temperature variations through the pulsation cycle, synthetic spectra
were fit to the ground-based {\it BV} photometry of Kilkenny 
et~al. (\cite{kk99}) and our {\it HST} ultraviolet spectrophotometry.  Before
the fitting, the spectra were binned in wavelength.  No information is lost
through the binning since the fits are to the shape of the spectrum, not to
individual lines. The spectra are noisy at short wavelengths, reflecting a
drop in detector sensitivity.  To avoid possible problems caused by this
increased noise, only the part of the spectra with
$\lambda \geq 1270$~\AA\ was used.

The pulsation
period was divided into phase bins and the spectra and photometry within each
bin were averaged.  No significant difference in the results was found with
the period divided into 15, 25, 50, or 99 bins.  We will report the results
found for 25 bins.

A grid of line-blanketed model atmospheres was calculated under the
assumption of plane-parallel geometry, hydrostatic equilibrium
and local thermodynamic equilibrium using the code {\sc sterne} described by
Jeffery \& Heber (\cite{j92}) and by Drilling et al. (\cite{d98}).
Following the latter, we assumed a composition for \object{LSS~3184} given
by  $n_{\rm H}=0, n_{\rm He}=0.99$ and $n_{\rm C}=0.003$,
where $n$ represents fractional
abundance by number, and all other elements were assumed to have
solar-like relative abundances. The grid extended between 15\,000 and 30\,000
K, $\log g=3.00$ and 4.00 (cgs units).
As will be shown later, changing the $\log g$ used in the model atmospheres
makes only a minor difference in the temperatures and angular radii derived.

\begin{figure}
\resizebox{\hsize}{!}{\rotatebox{270}{\includegraphics{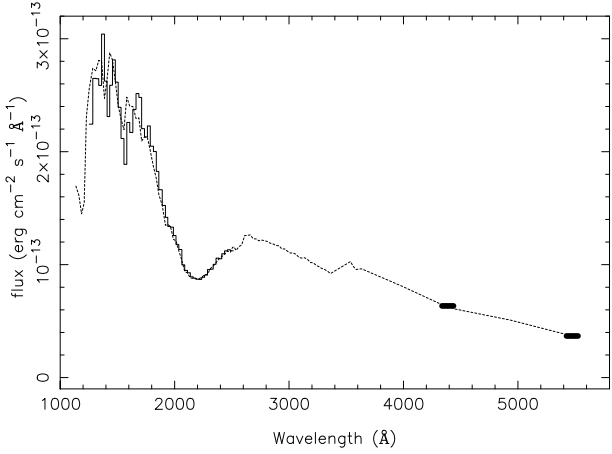}}}
\caption{Best fit synthetic spectrum (dashed curve) plotted with ultraviolet
spectrum (histogram) and {\it BV} photometry values.  This is for the phase bin
centered at 0.580.}
\label{figb}
\end{figure}

In the fitting procedure $T_{\rm eff}$, angular radius ($\alpha$), and
$E_{\rm B - V}$ were allowed to vary and the downhill simplex program
{\sc amoeba} (Press et~al. \cite{pt92}) was used to find the minimum $\chi ^2$
difference between the synthetic spectrum and the observed spectral and
photometric data at each phase (Jeffery et~al. \cite{j00}).
An example of the fit is shown in Fig.~\ref{figb}.

When the extinction $E_{\rm B - V}$ was allowed to vary we found the mean to be
$E_{\rm B - V} = 0.239 \pm 0.008$.
Because the extinction is not expected to vary with pulsation
phase, we chose to use $E_{\rm B - V} = 0.24$ throughout the cycle and did
a second iteration allowing only temperature and $\alpha$ to vary.

\subsection{Radius determination}
In determining the radius of \object{LSS~3184}, we make two assumptions.
First, we assume that the temperature (and thus
$\alpha$) and radial velocity (and thus $\Delta R_\star $) were measured at
approximately the same layer in the stellar atmosphere.  Second, because we
measure $\alpha$ perpendicular to the line of sight and $\Delta R_\star $
parallel to the line of sight, we are assuming that the pulsation is spherically
symmetric.  Thus we get $\Delta R_\star / R_\star =\Delta \alpha / \alpha$.
With these assumptions we can use a modified Baade's method (Baade \cite{b26})
to find $R_\star$, using
our previously determined values of $\alpha$ and $\Delta R_\star$.

There are two ways we can do this.  In the first, we choose two phase bins,
determine $\Delta \alpha / \alpha$ and $\Delta R_\star$ between them
and find $R_\star = \Delta R_\star \alpha / \Delta \alpha$, with
$\alpha$ and $R_\star$ defined at one of the chosen phase bins.  This method has
the disadvantage of using data from only two phase bins and thus ignoring
additional information available from the rest of the pulsation cycle.
In the second method we plot
$\Delta \alpha / \alpha$ versus $\Delta R_\star$.  The slope
of a linear fit to the data points is then $1 / R_\star$.  We use the second
method.

\section{Results and discussion}
The velocities, effective temperatures, and angular radii measured through the
pulsation cycle of \object{LSS~3184} are listed in Table~1.  The velocities are not
corrected for projection effects and are reported for the center of the phase
bins.  Temperatures and angular radii are reported for the beginning of the
phase bins. Data are missing where {\it HST} observations did not fall in the
affected phase bins.  Typical uncertainties are indicated.
\begin{table}
  \begin{minipage}{70mm}
   \caption{Velocity, temperature, and angular radius through the pulsation
cycle of \object{LSS~3184}. Velocity is not corrected for projection effects.}
   \begin{tabular}{@{}lccc@{}}
$\phi$ bin & $V\footnote[1]{center of phase bin}$ &
$T_{\rm eff}$\footnote[2]{beginning of phase bin} &
$\alpha^b$ \\
range & (${\rm km~s^{-1}}$) & (K) & ($10^{-11}$ arcsec) \\
 & $\pm 0.40$ & $\pm 90.$ & $\pm 0.005$ \\ \hline
0.00--0.04 &  18.58 & 22500 & 1.799   \\
0.04--0.08 &  14.80 & 22480 & 1.799   \\
0.08--0.12 &   3.39 & 22480 & 1.794   \\
0.12--0.16 &  -9.42 & 22450 & 1.790   \\
0.16--0.20 & -16.00 & 22350 & 1.793   \\
0.20--0.24 & -18.31 & 22230 & 1.796   \\
0.24--0.28 & -19.18 & 22040 & 1.807   \\
0.28--0.32 & -18.56 &  & \\
0.32--0.36 & -17.53 & 21700 & 1.831   \\
0.36--0.40 & -15.04 & 21500 & 1.846   \\
0.40--0.44 & -12.94 & 21340 & 1.857   \\
0.44--0.48 & -10.36 & 21180 & 1.867   \\
0.48--0.52 &  -7.59 & 21100 & 1.868   \\
0.52--0.56 &  -5.38 & 21000 & 1.874   \\
0.56--0.60 &  -2.82 & 20940 & 1.877   \\
0.60--0.64 &  -0.19 & 20930 & 1.876   \\
0.64--0.68 &   2.36 & 20900 & 1.877   \\
0.68--0.72 &   5.39 & 20900 & 1.877   \\
0.72--0.76 &   7.47 & 20990 & 1.870   \\
0.76--0.80 &  11.07 & 21100 & 1.864   \\
0.80--0.84 &  13.85 & 21330 & 1.849   \\
0.84--0.88 &  16.91 & 21540 & 1.840   \\
0.88--0.92 &  18.83 & 21750 & 1.832   \\
0.92--0.96 &  19.86 &  & \\
0.96--1.00 &  20.81 & 22410 & 1.802   \\ 
\end{tabular}
 \end{minipage}
\end{table}

\begin{figure}
\resizebox{\hsize}{!}{\includegraphics{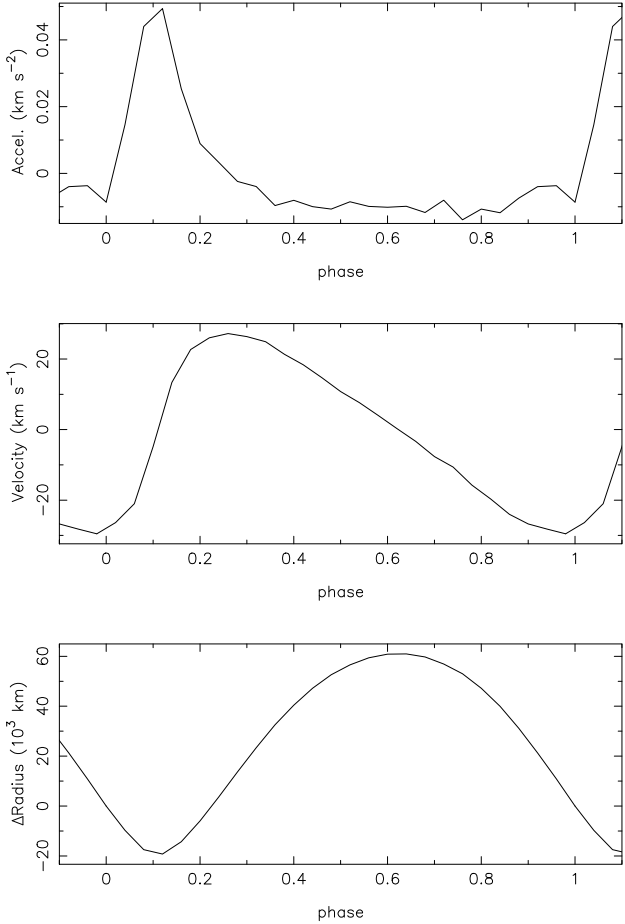}}
\caption{Surface acceleration, surface velocity, and change in radius of
\object{LSS~3184} through its pulsation cycle.}
\label{figc}
\end{figure}

The surface acceleration, surface velocity, and change in radius determined
for \object{LSS~3184} using the AAT spectra are shown in Fig.~\ref{figc}.  Note that
the surface velocity reported here is the velocity measured from the spectra
multiplied by $-1.42$ to correct for projection
effects and make positive velocity be away from the star's center.
The change in radius has been set so that it is zero at photometric
maximum, $\phi = 0$.

\begin{figure}
\resizebox{\hsize}{!}{\includegraphics{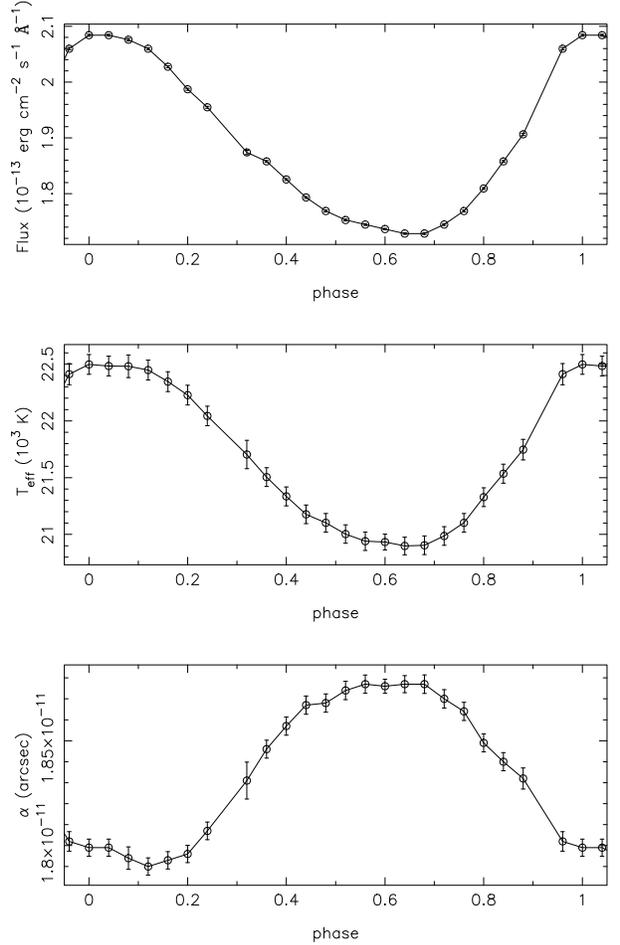}}
\caption{Flux $\langle F_{1270-2508} \rangle $,
effective temperature, and angular radius of \object{LSS~3184} through its
pulsation cycle. The curves simply connect the data points and are not fits
to the data.}
\label{figd}
\end{figure}

Fig.~\ref{figd} shows the integrated
flux between 1270 and 2508~\AA\ and the temperature
and angular radius determined by fitting synthetic spectra to {\it HST} UV
spectra and ground-based {\it BV} photometry through the pulsation cycle.

\begin{figure}
\resizebox{\hsize}{!}{\rotatebox{270}{\includegraphics{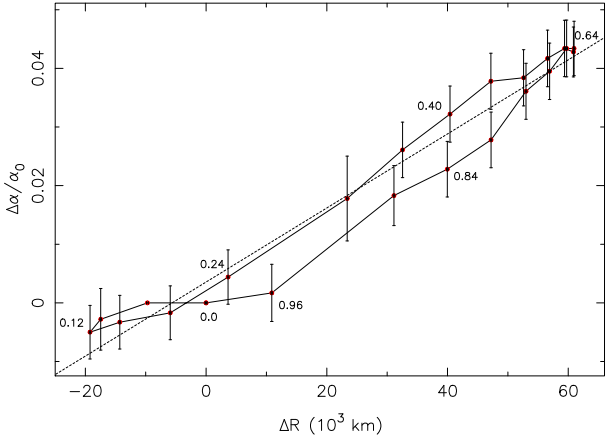}}}
\caption{$\Delta \alpha / \alpha_0$ versus $\Delta R_\star$ through
the pulsation cycle of \object{LSS~3184}.
Points are connected in order of phase with
$\phi = 0$ at the origin.  The dashed line is a linear least squares fit to all
data points. Error bars for $\Delta R_\star$ are smaller than the symbols.
Numbers next to symbols indicate the corresponding pulsation phase.}
\label{fige}
\end{figure}

$\Delta \alpha / \alpha_0$ is plotted against $\Delta R_\star$ in
Fig.~\ref{fige}.  The `0' subscript indicates that the reference bin is at
$\phi = 0$.
\begin{figure}
\resizebox{\hsize}{!}{\rotatebox{270}{\includegraphics{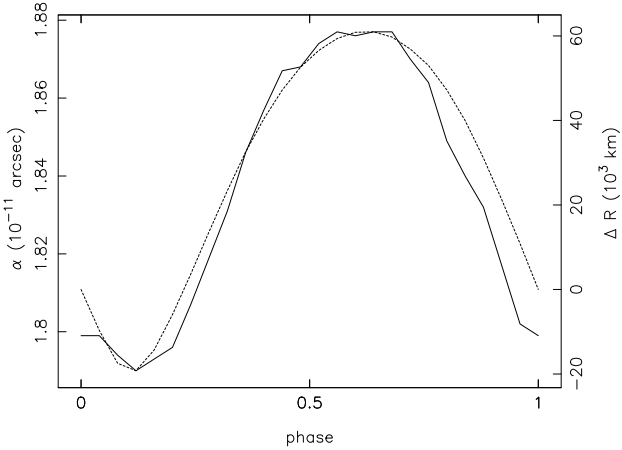}}}
\caption{$\alpha$ (solid curve) and $\Delta R$ (dashed curve) versus
pulsation phase.}
\label{figf}
\end{figure}
As is seen in Figures \ref{figc} and \ref{figd}, the shapes of
the $\alpha$ and $\Delta R_\star$ curves are not identical, which is the cause
of the non-linear, looped shape of the $\Delta \alpha / \alpha$ versus
$\Delta R_\star$ curve.  If the $\alpha$ and $\Delta R_\star$ curves are
normalized and placed on the same plot (Fig.~\ref{figf}), the differences in the
curves are more noticeable.  Figure 12 from Kilkenny et~al. (\cite{kk99}) shows
a similar loop on the right hand side of the angular radius versus stellar
linear radius plot.  Our work cannot be used as an independent confirmation of
the non-linear angular versus stellar radius curve, however, as we have used the
same {\it BV} photometry as Kilkenny et~al.

The non-linear shape makes determining the radius more difficult.
It raises questions about the assumption made in
using Baade's method to find $R_\star$, that $\Delta R_\star$ and
$\Delta \alpha$ are measurements of the same quantity, with $\Delta \alpha$
decreased by a factor proportional to the distance to \object{LSS~3184}.
One possible explanation of the discrepancy is that we are measuring
$\Delta R_\star$ and $\alpha$ at different layers in the atmosphere.
$\Delta R_\star$ is measured using optical spectra, while $\alpha$ is measured
using ultraviolet spectra and optical photometry. It is possible that the
layer where the optical lines are formed expands and contracts a bit differently
than the layer where the ultraviolet continuum is formed. It is also possible
that the star and/or its pulsations are not spherically symmetric, as might
occur if the star were flattened by rotation, so that the
measured $\Delta \alpha$, which is perpendicular to the line of sight, and
$\Delta R_\star$, which is parallel to the line of sight, act
differently.  Further, the atmosphere is not
static. The atmosphere's temperature is constantly changing, and the pulsating
layers undergo a substantial compression at minimum radius,
as shown by the spike in the acceleration in Fig~\ref{figc}, which may cause
nonadiabatic effects.  So the temperature may be acting
differently in the expanding part of the phase than in the contracting part.

However, it is encouraging that the slopes of the expanding (the upper half of
the loop in Fig.~\ref{fige}) and contracting (lower half of the loop) parts of
the $\Delta \alpha / \alpha$ versus $\Delta R_\star$ curve are not
too different.  {$\Delta \alpha$ and $\Delta R_\star$ are still
correlated.

If we find the slope of the curve using a least squares fit to all of
the data points (dashed line in
Fig.~\ref{fige}) then the radius at photometric maximum ($\phi = 0$),
the inverse of the slope, is $2.27 \pm 0.10 R_\odot$, where the uncertainty
is derived from the standard error in the least squares fit to the slope.
The average $\Delta R_\star$ over the pulsation cycle, with
$\Delta R_\star \equiv 0$ at $\phi = 0$, is 28\,190~km, or $0.04 R_\odot$.
So if we use all the $\Delta \alpha / \alpha_0$
versus $\Delta R_\star$ data points, we find the mean $R_\star$ to be
$\langle R_\star \rangle = 2.31 R_\odot$.  This is larger than the
$1.35 R_\odot$ mean radius found by
Kilkenny et~al. (\cite{kk99}) for \object{LSS~3184} and is closer to
the $1.91 R_\odot$ mean radius found for \object{V652~Her} by Lynas-Gray et~al.
(\cite{l84}).

We tested the effects of using only
a portion of the $\Delta \alpha / \alpha$ versus
$\Delta R_\star$ curve to determine $R_\star$, though we have no reason to
reject any particular data points.
If we use the points on the upper part of the curve, with phase
${\rm 0.16 \leq \phi \leq 0.56 }$, then we find
$\langle R_\star \rangle = 2.16 \pm 0.07 R_\odot$. If
we use the points on the lower part of the curve, ignoring the points to the
left of the change in slope at $\Delta R_\star \approx 10\,000$~km, so that
${\rm 0.60 \leq \phi \leq 0.96 }$, we find
$\langle R_\star \rangle = 1.75 \pm 0.08 R_\odot$. If we use
only the points to the right of $\Delta R_\star = 0$~km then we find
$\langle R_\star \rangle = 1.97 \pm 0.10 R_\odot$.
In all cases, even the extreme one where we
reject all points but those on the lower part of the curve, our
$\langle R_\star \rangle $ is
larger than $R_\star = 1.35 \pm 0.15 R_\odot$, the value
found by Kilkenny et~al. (\cite{kk99}).

Drilling et~al. (\cite{d98}) estimated
$\log g = 3.35 \pm 0.1$ for \object{LSS~3184}.
If we use this with our $\langle R_\star \rangle = 2.31 R_\odot$ in the formula
$g = G M R^{-2}$, we find $M_\star = 0.42 \pm 0.12 M_\odot$.
This is larger than $0.15 M_\odot$,
the mass found by Kilkenny et~al. (\cite{kk99}) for \object{LSS~3184} and
closer to $0.7^{+0.4}_{-0.3}$, the estimated mass of \object{V652~Her}
(Lynas-Gray et~al. \cite{l84}).  Our larger estimate results
mainly from the larger peak-to-peak range of radial velocities we measured.
It is likely that the smaller velocity amplitude measured by Kilkenny et~al.
resulted from a combination of the lower spectral resolution and the lower
signal to noise in the spectra they used.  This meant that even the best
cross correlation template still contained some velocity broadening, thus
diluting the velocity amplitude.

There are several sources of uncertainty in our mass determination.
Uncertainty in the $\log g$ used is a major contributor.
We note that it is impossible to determine the
pulsation phase of \object{LSS~3184}
when the spectrum used in the analysis of Drilling et~al. (\cite{d98}) was taken
in 1985, as $\dot{P}$ is unknown.
They estimated the temperature of \object{LSS~3184} at $23\,300 \pm 700$~K.
Our data yield a similar value, $\langle T_{\rm eff} \rangle =23\,230$~K,
if we use $E_{\rm B - V} = 0.27$, the
extinction they used.  However, because temperature, line strengths,
and other parameters presumably varied throughout the 1-hour exposure for their
spectrum, it is possible that the errors in their temperature and gravity
determinations are larger than the formal errors quoted.
If we assume that \object{LSS~3184} has the same $\log g$
as \object{V652~Her}, $\log g = 3.68$
(Jeffery et~al. \cite{j99}), instead of $\log g = 3.35$, then we find
that it has $M_\star = 0.92 M_\odot$ 

As mentioned earlier, the model atmospheres used to calculate temperature and
angular radius from {\it HST} UV spectra and ground-based
{\it BV} photometry assumed $\log g = 3.50$.  The assumed gravity
enters into the stellar parameter calculations in the stellar atmospheres used
and in calculating the mass from the radius.  Changing $\log g$ in the
model atmospheres by 0.50~dex (to 3.00 or 4.00), but using $\log g = 3.35$
to calculate the mass as before, changes the
$\langle T_{\rm eff} \rangle$, $\langle \alpha \rangle$, and
$\langle R_\star \rangle$ derived by about 1 per cent, and thus has only a small
effect ($\sim 3$ per cent) on the mass derived.
The mass finally derived is proportional to $g$.

Uncertainty in the extinction also adds uncertainty to the stellar
parameters calculated.  Using $E_{\rm B - V} = 0.25$  instead of 0.24 increases
the derived $\langle T_{\rm eff} \rangle$ by 510~K, or about 2 per cent,
but changes $\langle \alpha \rangle$ by only 0.2 per cent, and thus has a very
minor effect on the derived stellar radius and mass.

The derived radius varies proportionally with the projection factor used to
transform the measured stellar radial velocities into surface velocities.
For example, using a projection factor of 1.31, as Lynas-Gray et~al.
(\cite{l84}) used in their analysis of \object{V652~Her}, instead of 1.42
would give $\langle R_\star \rangle = 2.12 R_\odot$ instead of
$\langle R_\star \rangle = 2.31 R_\odot$.  The smaller radius would give
$M_\star = 0.37 M_\odot$ instead of $M_\star = 0.42 M_\odot$.

\section{Conclusions}

In this paper we report new determinations of the radius
($\langle R_\star \rangle = 2.31 \pm 0.10 R_\odot$) and
mass ($M_\star = 0.42 \pm 0.12 M_\odot$)
of the pulsating hydrogen-deficient star \object{LSS~3184}.  The radial velocity data and
temperature measurements are more reliable than those used for previous radius
and mass determinations for this star, so our estimates are likely to be
closer to the star's actual parameters.  Further improvements can be
accomplished by improving the $\log g$ and chemical composition estimates for
\object{LSS~3184}.
In addition, there needs to be further study of the difference
between the shapes of the angular radius and physical radius curves based on
temperature and flux measurements, and radial velocity measurements,
respectively.  It is important to determine if the difference in shapes is a
sign that some of our data or a step in our analysis is flawed, or if there is
something happening in the star itself that causes the two measures of radius
to behave differently with pulsational phase.

%---------------------------------------------------------------
\begin{acknowledgements}
We thank Dr. D. Kilkenny for providing the differential photometric data used
in the temperature analysis. We thank the referee, Dr. D. Kurtz, for his
helpful comments.  We acknowledge financial support from the former
Department of Education of Northern Ireland and the UK PPARC (grant Ref
PPA/G/S/1998/00019). 
\end{acknowledgements}

\end{document}